\newif\ifGALLEYversion\GALLEYversionfalse
\documentclass[prb,twocolumn,showpacs,amsmath,amssymb,citeautoscript]{revtex4}

\usepackage{ifthen}
\usepackage{graphicx}
\usepackage{dcolumn}

\ifthenelse{\boolean{GALLEYversion}}{
    \marginparwidth 2.7in
    \marginparsep 0.5in
    \def\dvm#1{\marginpar{\small DV: #1}}
    \def\rem#1{\marginpar{\small RE: #1}}
}{
    \def\dvm#1{\marginpar{\small NOTE}}
    \def\rem#1{\marginpar{\small NOTE}}
}

\begin{document}
\title{{\it Ab initio}
calculations of BaTiO$_3$ and  PbTiO$_3$
(001) and (011) surface structures
}
\author{R. I. Eglitis, and
David Vanderbilt}
\affiliation{Department of Physics and Astronomy,
Rutgers University, 136 Frelinghuysen Road, Piscataway,
New Jersey 08854-8019, USA}

\date{October 3, 2007}

\begin{abstract}
We present and discuss the results of calculations of surface
relaxations and rumplings for the (001) and (011) surfaces of
BaTiO$_3$ and PbTiO$_3$, using a hybrid B3PW description of
exchange and correlation.  On the (001) surfaces, we consider both
AO (A = Ba or Pb) and TiO$_2$ terminations.  In the former
case, the surface AO layer is found to relax inward for both
materials, while outward relaxations of all atoms in the second
layer are found at both kinds of (001) terminations and for
both materials.  The surface relaxation energies of BaO and
TiO$_2$ terminations on BaTiO$_3$ (001) are found to be comparable,
as are those of PbO and TiO$_2$ on PbTiO$_3$ (001), although in
both cases the relaxation energy is slightly larger for the TiO$_2$
termination.  As for the (011) surfaces, we consider three types
of surfaces, terminating on a TiO layer, a Ba or Pb layer, or an
O layer.  Here, the relaxation energies are much larger for the
TiO-terminated than for the Ba or Pb-terminated surfaces.  The
relaxed surface energy for the O-terminated surface is about the
same as the corresponding average of the TiO and Pb-terminated
surfaces on PbTiO$_3$, but much less than the average of the TiO
and Ba-terminated surfaces on BaTiO$_3$.  We predict a considerable
increase of the Ti-O chemical bond covalency near the BaTiO$_3$
and PbTiO$_3$ (011) surface as compared to both the bulk and
the (001) surface.
\end{abstract}


\maketitle
\section{INTRODUCTION}

Thin films of ABO$_3$ perovskite ferroelectrics play an important role
in numerous microelectronic, catalytic, and other high-technology
applications, and are frequently used as substrates for growth of
other materials such as cuprate superconductors \cite{scott,lin}.
Therefore, it is not surprising that a large number of
{\it ab initio} quantum mechanical calculations
\cite{nogu,pad,mey,cor,fu,chen,meyer, bung,krc,umen,lai}, as well as
several classical shell-model (SM) studies \cite{heif,tin,heif1},
have dealt with the atomic and electronic structure of the
(001) surface of  BaTiO$_3$, PbTiO$_3$, and SrTiO$_3$ crystals.
In order to study the dependence of the surface relaxation properties
on the exchange-correlation functionals and the type of basis (localized
vs.~plane-wave) used in the calculations, a detailed comparative study
of SrTiO$_3$ (001) surfaces based on ten different quantum-mechanical
techniques \cite{heif2,heif3} was recently performed.

Due to intensive development and progressive miniaturization of electronic
devices, the surface structure as well as the electronic properties
of the ABO$_3$ perovskite thin films have been extensively
studied experimentally during the last years. The SrTiO$_3$ (001) surface
structure has been analyzed by means of low-energy electron diffraction
(LEED) \cite{bic}, reflection high-energy electron diffraction (RHEED)
\cite{hik}, X-ray photoelectron spectroscopy (XPS), ultraviolet electron
spectroscopy (UPS), medium-energy ion scattering (MEIS) \cite{ike}, and
surface X-ray diffraction (SXRD) \cite{cha}. Nevertheless it is
important to note that the LEED \cite{bic} and RHEED \cite{hik}
experiments contradict each other in the sign (contraction or expansion)
of the interplanar distance between top metal atom and the second crystal
layer for the SrO-terminated SrTiO$_3$ (001)
surface. The most recent experimental studies on the SrTiO$_3$ surfaces
include a combination of XPS, LEED, and time-of-flight scattering and recoil
spectrometry (TOF-SARS) \cite{hei}, as well as metastable impact electron
spectroscopy (MIES) \cite{mau}. In these recent studies, well-resolved
1$\times$1 LEED patterns were obtained for the TiO$_2$-terminated SrTiO$_3$
(001) surface. Simulations of the TOF-SARS azimuthal scans indicate that the
O atoms are situated 0.1\,\AA\ above the Ti layer (surface plane) in the
case of the TiO$_2$-terminated SrTiO$_3$ (001) surface.

While the (001) surfaces of SrTiO$_3$, BaTiO$_3$ and PbTiO$_3$ have been
extensively studied, much less is known about the (011) surfaces.  The
scarcity of information about these surfaces is likely due to the polar
character of the (011) orientation. (011) terminations of SrTiO$_3$ have
frequently been observed, but efforts towards the precise characterization
of their atomic-scale structure and corresponding electronic properties
has only begun in the last decade, specifically using atomic-force
microscopy \cite{szo}, scanning tunneling microscopy (STM), Auger
spectroscopy, and low-energy electron-diffraction (LEED) \cite{bru}
methods.

To the best of our knowledge, very few {\it ab initio} studies of
perovskite (011) surfaces exist.  The first {\it ab initio} study
of the electronic and atomic structures of several (1$\times$1)
terminations of the (011) polar orientation of the SrTiO$_3$
surface was performed by Bottin {\it et al.} \cite{bott}
One year later Heifets {\it et al.} \cite{heifhf} performed very
comprehensive {\it ab initio} Hartree-Fock calculations for four
possible terminations (TiO, Sr, and two kinds of O terminations)
of the SrTiO$_3$ (011) surface.  Recently Heifets {\it et al.}
\cite{heifbz} performed {\it ab initio} density-functional
calculations of the atomic structure and charge redistribution
for different terminations of the BaZrO$_3$ (011) surfaces.
However, despite the high technological potential of BaTiO$_3$ and
PbTiO$_3$, we are unaware of any previous
{\it ab initio} calculations performed for the BaTiO$_3$ and
PbTiO$_3$ (011) surfaces. In this study, therefore, we have investigated
the (011) as well as the (001) surfaces of BaTiO$_3$ and PbTiO$_3$,
with an emphasis on the effect of the surface relaxation and rumpling,
surface energies, and the charge redistributions and changes in
bond strength that occur at the surface.

\section{Preliminaries}

\subsection{Computational method}

We carry out first-principles calculations in the framework of
density-functional theory (DFT) using the CRYSTAL computer code
\cite{saun}. Unlike the plane-wave codes employed in many previous
studies \cite{coh,coh1}, CRYSTAL uses localized Gaussian-type basis
sets.  In our calculations, we adopted the basis sets developed for
BaTiO$_3$ and PbTiO$_3$ in Ref.~[\onlinecite{pis}].
Our calculations were performed using the hybrid exchange-correlation 
B3PW functional involving a mixture of non-local Fock exact exchange,
LDA exchange, and Becke's gradient corrected exchange functional \cite{bec}, 
combined with the non-local gradient corrected correlation 
potential of Perdew and Wang \cite{per1,per2,per3}.
We chose the hybrid B3PW functional for our current study because it
yields excellent results for the SrTiO$_3$, BaTiO$_3$, and PbTiO$_3$
bulk lattice constant and bulk modulus \cite{heif2,pis}.

The reciprocal-space integration was performed by
sampling the Brillouin zone with an
8$\times$8$\times$8 Pack-Monkhorst mesh \cite{monk}, which provides a balanced
summation in direct and reciprocal spaces. To achieve high accuracy, large
enough tolerances of 7, 8, 7, 7, and 14
were chosen for the dimensionless Coulomb overlap,
Coulomb penetration, exchange overlap, first exchange pseudo-overlap,
and second exchange pseudo-overlap parameters, respectively \cite{saun}.

An advantage of the CRYSTAL code is
that it treats isolated 2D slabs, without any artificial periodicity in the $z$
direction perpendicular to the surface, as commonly employed in most previous
surface band-structure calculations (e.g., Ref.~[\onlinecite{chen}]).
In the present
{\it ab initio} investigation, we have studied several isolated periodic
two-dimensional slabs of cubic BaTiO$_3$ and PbTiO$_3$ crystals
containing 7 planes of atoms.

\subsection{Surface geometries}

\begin{figure}
\includegraphics[width=2.6in]{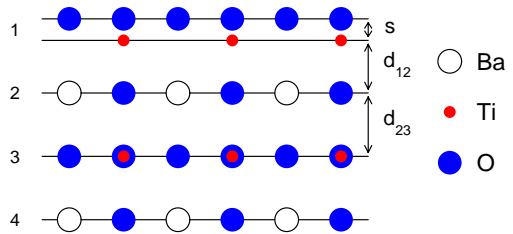}
\caption{(Color online.)
Side view of a TiO$_2$-terminated BaTiO$_3$ (001) surface
with the definitions of the surface rumpling $s$ and the near-surface
interplanar distances $d$$_{12}$ and $d$$_{23}$, respectively.}
\end{figure}

The BaTiO$_3$ and PbTiO$_3$ (001) surfaces were modeled using symmetric
(with respect to the mirror plane) slabs consisting of seven alternating
TiO$_2$ and BaO or PbO layers, respectively. One of these slabs was
terminated by BaO planes for the BaTiO$_3$ crystal (or PbO planes for
PbTiO$_3$) and consisted of a supercell containing 17 atoms.
The second slab was terminated by TiO$_2$ planes for both
materials and consisted of a supercell containing 18 atoms.
These slabs are non-stoichiometric, with unit cell formulae
Ba$_4$Ti$_3$O$_{10}$ or Pb$_4$Ti$_3$O$_{10}$, and Ba$_3$Ti$_4$O$_{11}$
or Pb$_3$Ti$_4$O$_{11}$ for BaTiO$_3$ and PbTiO$_3$ perovskites,
respectively. These two (BaO or PbO and TiO$_2$) terminations are
the only two possible flat and dense (001) surfaces for the BaTiO$_3$
or PbTiO$_3$ perovskite lattice structure. The
sequence of layers at the TiO$_2$-terminated (001) surface of
BaTiO$_3$ is illustrated in Fig.~1.

Unlike the (001) cleavage of BaTiO$_3$ or PbTiO$_3$, which naturally
gives rise to non-polar BaO (or PbO) and TiO$_2$ terminations, a naive
cleavage of BaTiO$_3$ or PbTiO$_3$ to create (011) surfaces leads to the
formation of polar surfaces.  For example, the stacking of the BaTiO$_3$
crystal along the (011) direction consists of alternating planes of O$_2$
and BaTiO units having nominal charges of $-4e$ and $+4e$ respectively,
assuming O$^{2-}$, Ti$^{4+}$, and Ba$^{2+}$ constituents.  (Henceforth
we shall use BaTiO$_3$ for presentation purposes, but everything that
is said will apply equally to the PbTiO$_3$ case.) Thus,
a simple cleavage leads to O$_2$-terminated and BaTiO-terminated (011)
surfaces that are {\it polar} and have nominal surface charges of $-2\,e$
and $+2\,e$ per surface cell respectively.  These are shown as the top
and bottom surfaces in Fig.~2(a) respectively.  If uncompensated, the
surface charge would lead to an infinite electrostatic cleavage energy.
In reality, the polar surfaces would probably become metallic in order
to remain neutral, but in view of the large electronic gaps in the
perovskites, such metallic surfaces would presumably be unfavorable.
Thus, we may expect rather generally that such polar crystal terminations
are relatively unstable in this class of materials \cite{nogu}.

\begin{figure}
\includegraphics[width=3.0in]{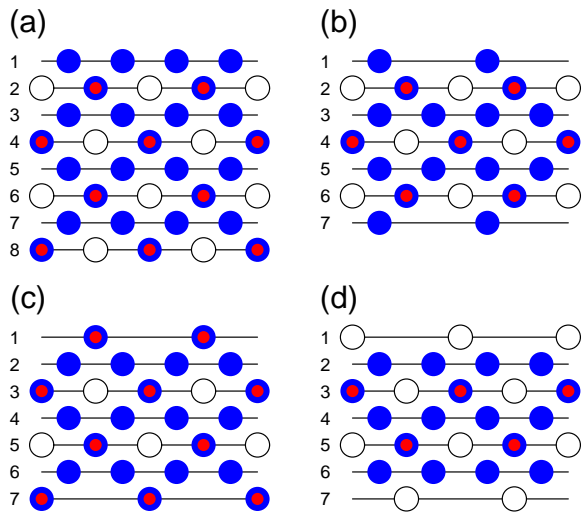}
\caption{(Color online.)
Side views of slab geometries used to study BaTiO$_3$ (011) surfaces.
(a) Stoichiometric 8-layer slab with O$_2$-terminated and
BaTiO-terminated surfaces at top and bottom respectively.
(b) 7-layer slab with O-terminated surfaces.
(c) 7-layer slab with TiO-terminated surfaces.
(d) 7-layer slab with Ba-terminated surfaces.}
\end{figure}

On the other hand, if the cleavage occurs in such a way as to
leave a half layer of O$_2$ units on each surface, we obtain
the non-polar surface structure shown in Fig.~2(b).  Every other
surface O atom has been removed, and the remaining O atoms occupy the same
sites as in the bulk structure.  We shall refer to this as the
``O-terminated'' (011) surface, in distinction to the
``O$_2$-terminated'' polar surface already discussed in Fig.~2(a).
The non-polar nature of the O-terminated surface can be confirmed by noting
that the 7-layer 15-atom Ba$_3$Ti$_3$O$_9$ slab shown in Fig.~2(b), which
has two O-terminated surfaces, is neutral.  It is also possible
to make non-polar TiO-terminated and Ba-terminated surfaces, as shown
in Figs.~2(c) and (d), respectively.  This is accomplished by splitting
a BaTiO layer during cleavage, instead of splitting an O$_2$ layer.
For the TiO and Ba-terminated surfaces, we use 7-layer slabs having
composition Ba$_2$Ti$_4$O$_{10}$ (16 atoms) and Ba$_4$Ti$_2$O$_8$ (14 atoms)
as shown in Fig.~2(c-d), respectively.  These are again neutral,
showing that the surfaces are non-polar (even though they no
longer have precisely the bulk BaTiO$_3$ stoichiometry).

\section{RESULTS OF CALCULATIONS}

\subsection{BaTiO$_3$ and PbTiO$_3$ (001) surface structure}

\begin{table*}
\caption{Vertical atomic relaxations (in percent of bulk lattice constant)
for BaTiO$_3$ and PbTiO$_3$ (001) surfaces.
Positive sign corresponds to outward atomic displacement. 
`SM' indicates shell-model calculation of Ref.~[\onlinecite{heif1}];
`LDA' are previous calculations of  Ref.~[\onlinecite{pad}]
and [\onlinecite{mey}] for BaTiO$_3$ and PbTiO$_3$ respectively.}
\begin{ruledtabular}
\begin{tabular}{cccdddcccdd}
\multicolumn{6}{c}{BaTiO$_3$ (001) surface relaxations} &
   \multicolumn{5}{c}{PbTiO$_3$ (001) surface relaxations} \\
Termination & Layer & Ion &
   \multicolumn{1}{c}{This study} &
   \multicolumn{1}{c}{SM} & 
   \multicolumn{1}{c}{LDA} &
   Termination & Layer & Ion & \multicolumn{1}{c}{This study} &
   \multicolumn{1}{c}{LDA} \\
\hline
BaO    & 1 & Ba &-1.99 &-3.72 &-2.79 & PbO   & 1 &Pb &-3.82 &-4.36 \\
       &   & O  &-0.63 & 1.00 &-1.40 &       &   & O &-0.31&-0.46 \\
       & 2 & Ti & 1.74 & 1.25 & 0.92 &       & 2 &Ti & 3.07& 2.39  \\
       &   & O  & 1.40 & 0.76 & 0.48 &       &   & O & 2.30& 1.21 \\
       & 3 & Ba &      &-0.51 & 0.53 &       & 3 & Pb&    &-1.37 \\
       &   & O  &      & 0.16 & 0.26 &       &   & O &    &-0.20 \\
TiO$_2$& 1 & Ti &-3.08 &-2.72 &-3.89 &TiO$_2$& 1 &Ti &-2.81&-3.40 \\
       &   & O  &-0.35 &-0.94 &-1.63 &       &   &O  & 0.31&-0.34 \\
       & 2 & Ba & 2.51 & 2.19 & 1.31 &       & 2 &Pb & 5.32& 4.53 \\
       &   & O  & 0.38 &-0.17 &-0.62 &       &   &O  & 1.28& 0.43 \\
       & 3 & Ti &      &-0.33 &-0.75 &       & 3 &Ti &     &-0.92 \\
       &   & O  &      &-0.01 &-0.35 &       &   &O  &     &-0.27 \\
\end{tabular}
\end{ruledtabular}
\end{table*}

In the present calculations of the BaTiO$_3$ and PbTiO$_3$ (001) surface
atomic structure, we allowed the atoms located in the two outermost surface
layers to relax along the $z$-axis (the forces along the $x$ and
$y$-axes are zero by symmetry).  Here we use the term ``layer'' to refer
to a BaO, PbO, or TiO$_2$ plane, so that there are two layers per stacked
unit cell.  For example, on the BaO or PbO-terminated surfaces, the top
layer is BaO or PbO and the second layer is TiO$_2$; displacements of
the third-layer atoms were found to be negligibly small in our calculations
and thus were neglected.

The calculated atomic displacements for the TiO$_2$ and
BaO-terminated (001) surfaces of BaTiO$_3$, and for the TiO$_2$ and
PbO-terminated (001) surfaces of PbTiO$_3$, are presented in Table I.
For BaTiO$_3$ (001), comparisons are also provided with the
surface atomic displacements obtained by Padilla and
Vanderbilt \cite{pad} using plane-wave DFT methods in the local-density
approximation (LDA), and by Heifets {\it et al.} using a classical
shell-model (SM) approach \cite{heif1}.  Similarly, for PbTiO$_3$ (001),
Table I shows comparisons with the plane-wave LDA calculations of
Meyer and Vanderbilt \cite{mey}.  The relaxation of the surface metal
atoms in both the BaTiO$_3$ and PbTiO$_3$ surfaces
is much larger than that of the oxygen ions, leading to a considerable
surface rumpling, which we quantify via a parameter $s$ defined as
the relative displacement of the oxygen with respect to the metal atom
in a given layer.
The surface rumpling and relative displacements of three
near-surface planes are presented in Table II. According to our calculations,
atoms of the first surface layer relax inwards (i.e., towards the bulk)
for BaO and PbO terminations of both materials. Our calculations are in a
qualitative agreement with the {\it ab initio} calculations performed
by Padilla and Vanderbilt \cite{pad} for BaTiO$_3$, and by Meyer and
Vanderbilt \cite{mey} for PbTiO$_3$.
However, the predictions of the SM calculation disagree with the
first-principles calculations; the SM predicts that
the first-layer oxygen ions relax outward on the
BaO-terminated BaTiO$_3$ (001) surface\cite{heif1},
rather than inwards. However, the magnitudes of the displacements
are relatively small ($-$0.63\%  of the lattice constant $a_0$
in this study and 1.00\% of $a_0$ in the SM calculations \cite{heif1})
which may be close to the error bar of the classical shell model.
Outward relaxations of all
atoms in the second layer are found at both (001) terminations of
the BaTiO$_3$ and PbTiO$_3$ surfaces. From Table I, we can conclude
that the magnitudes of the atomic displacements calculated using
different {\it ab initio} methods and using the classical shell model
are in a reasonable agreement.

\begin{table}[b]
\caption{Calculated surface rumpling $s$ and interlayer
displacements $\Delta$$d$$_{ij}$ (in percent of bulk lattice
constant) for near-surface planes on the BaO or PbO and
TiO$_2$-terminated (001) surfaces of BaTiO$_3$ and PbTiO$_3$.}
\begin{ruledtabular}
\begin{tabular}{ldddddd}
 & \multicolumn{3}{c}{BaO or PbO term.} &
   \multicolumn{3}{c}{TiO$_2$ termination} \\
 & \multicolumn{1}{c}{$s$} & \multicolumn{1}{c}{$\Delta d_{12}$} &
   \multicolumn{1}{c}{$\Delta d_{23}$} & \multicolumn{1}{c}{$s$} &
   \multicolumn{1}{c}{$\Delta d_{12}$} & \multicolumn{1}{c}{$\Delta d_{23}$} \\
\hline
\multicolumn{7}{l}{BaTiO$_3$ (001)}\\
~~This study  & 1.37 & -3.74 & 1.74 & 2.73 & -5.59 & 2.51 \\
~~LDA, Ref.~\onlinecite{pad} & 1.39 & -3.71 & 0.39 & 2.26 & -5.20 & 2.06 \\
~~SM, Ref.~\onlinecite{heif1} & 4.72 & -4.97 & 1.76 & 1.78 & -4.91 & 2.52 \\
\multicolumn{7}{l}{PbTiO$_3$ (001)}\\
~~This study & 3.51 & -6.89 & 3.07 & 3.12 & -8.13 & 5.32 \\
~~LDA, Ref.~\onlinecite{mey} & 3.9 & -6.75 & 3.76 & 3.06 & -7.93 & 5.45 \\
\end{tabular}
\end{ruledtabular}
\end{table}

In order to compare the calculated surface structures further with
experimental results, the surface rumpling $s$ and the changes
in interlayer distances $\Delta d_{12}$ and $\Delta d_{23}$, as
defined in Fig.~1, are presented in Table II. Our calculations
of the interlayer distances are based on the positions of
relaxed metal ions, which are known to be much stronger electron
scatters that the oxygen ions \cite{bic}.

For BaTiO$_3$ (001), the rumpling of TiO$_2$-terminated surface
is predicted to exceed that of BaO-terminated surface
by a factor of two. This finding is in line with the values of
surface rumpling reported by Padilla and Vanderbilt \cite{pad}.
In contrast, PbTiO$_3$ demonstrates practically the same rumpling for both
terminations. From Table II one can see that qualitative agreement
between all theoretical methods is obtained. In particular, the
relaxed (001) surface structure shows a reduction of interlayer
distance $\Delta d_{12}$ and an expansion of $\Delta d_{23}$
according to all {\it ab initio} and shell-model results.

As for experimental confirmation of these results, we are
unfortunately unaware of experimental measurements of
$\Delta d_{12}$ and $\Delta d_{23}$ for the BaTiO$_3$ and
PbTiO$_3$ (001) surfaces.  Moreover, for the case of the
SrO-terminated SrTiO$_3$ (001) surface, existing
LEED \cite{bic} and RHEED \cite{hik} experiments actually
contradict each other regarding the sign of $\Delta d_{12}$.
In view of the absence of clear experimental determinations of
these parameters, therefore, the first-principles calculations
are a particularly important tool for understanding the surface
properties.

\subsection{BaTiO$_3$ and PbTiO$_3$ (011) surface structure}

\begin{table*}
\caption{Calculated surface relaxations of BaTiO$_3$ and PbTiO$_3$
(011) surfaces (in percent of the lattice constant) for the three
surface terminations. `SM' indicates comparative results from
the shell-model calculation of Ref.~[\onlinecite{heif1}].
}
\begin{ruledtabular}
\begin{tabular}{ccddddccdd}
\multicolumn{6}{c}{BaTiO$_3$ (011) surface} &
  \multicolumn{4}{c}{PbTiO$_3$ (011) surface} \\
Layer & Ion & \multicolumn{1}{c}{$\Delta z$} & \multicolumn{1}{c}{$\Delta y$} &
  \multicolumn{1}{c}{$\Delta z$ (SM)} & \multicolumn{1}{c}{$\Delta z$ (SM)} &
  Layer & Ion & \multicolumn{1}{c}{$\Delta z$} & \multicolumn{1}{c}{$\Delta y$} \\
\hline
\multicolumn{6}{l}{TiO-terminated} & \multicolumn{4}{l}{TiO-terminated} \\
1  & Ti & -7.86 &   & -6.93 &   & 1 & Ti & -8.13 &  \\
1  & O & 2.61 &   & 6.45 &   & 1 & O & 3.30  &  \\
2  & O  & -1.02 &   & -1.66 &   & 2 & O & -0.41 &  \\
3  & Ba & -0.88 &   & -3.85 &   & 3 & Pb & -2.54 &  \\
3  & O &   &   & -2.40 &   & 3 & O & -4.07 &  \\
3  & Ti &   &   & 1.59 &   & 3 & Ti & 0.30 &   \\
\multicolumn{6}{l}{Ba-terminated}  &  \multicolumn{4}{l}{Pb-terminated} \\
1 & Ba & -8.67 &   & -13.49 &   & 1 & Pb & -11.94 &  \\
2 & O & 0.80 &   & 2.80 &   & 2 & O & -0.61 &  \\
3 & Ti & 0.16 &   & -1.20 &   & 3 & Ti & 1.78 &  \\
3 & O & -0.43 &   & -2.94 &   & 3 & O & 1.67 &  \\
3 & Ba &   &   & 2.52 &   & 3 & Pb & 1.52 &  \\
\multicolumn{6}{l}{O-terminated}  &  \multicolumn{4}{l}{O-terminated} \\
1 & O & -5.40 & -1.67 & -11.16 & -6.70 & 1 & O & -7.37 & -0.07\\
2 & Ti & -0.15 & -6.38 & -1.83 & -5.33 & 2 & Ti & 0.20 & -2.54\\
2 & Ba & 1.54 & -1.27 & 4.84 & -2.21 & 2 & Pb & 0.18 & -7.50\\
2 & O & 1.95 & 2.97 & 4.54 & 5.90 & 2 & O & 0.51 & 2.19\\
3 & O & 0.90 & 4.49 & 6.52 & 5.58 & 3 & O & -0.41 & 3.30\\
\end{tabular}
\end{ruledtabular}
\end{table*}

To our knowledge, we have performed the first {\it ab initio}
calculations for BaTiO$_3$ and PbTiO$_3$ (011) surfaces.
We have studied the TiO$_2$-terminated, BaO or PbO-terminated,
and O-terminated surfaces illustrated in Fig.~2(c), (d), and (b),
respectively.  The computed surface atomic relaxations are reported
in Table III.

Focusing first on the BaTiO$_3$ surfaces, we find that the Ti
ions in the outermost layer of the TiO-terminated surface move
inwards (towards the bulk) by 0.0786\,$a_0$, whereas the O ions
in the outermost layer move outwards by a 0.0261\,$a_0$. The Ba
atoms in the top layer of the Ba-terminated surface of Fig.~2(d)
and the O atoms in the outermost layer of O-terminated surface
of Fig.~2(b) move inwards by 0.0867\,$a_0$ and 0.0540\,$a_0$,
respectively. The agreement between our {\it ab initio} B3PW and
the classical SM calculations is satisfactory for all three of
these surface terminations.  In particular, the directions of
the displacements of first and second-layer atoms coincide
for all three terminations.  This indicates that classical
SM calculations with a proper parameterization can serve as a
useful initial approximation for modeling the atomic structure
in perovskite thin films.

Turning now to our results for the  PbTiO$_3$ (011) surfaces,
we find that all metal atoms in the outermost layer move inwards
irrespective of the termination.  Surface oxygen atoms are
displaced outwards for the TiO-terminated surface, while oxygen
atoms move inwards in the O-terminated surface.  The displacement
patterns of atoms in the outermost surface layers are similar to
those of the BaTiO$_3$ (011) surfaces, as well as classical
shell model results for BaTiO$_3$ \cite{heif1}. For example,
the atomic displacement magnitudes of Ti and oxygen atoms in
the TiO-terminated PbTiO$_3$ (011) surface are $-$0.0813\,$a_0$
and 0.033\,$a_0$ respectively.  The Pb atom is displaced inwards
by 0.1194\,$a_0$ for the Pb-terminated surface, similar to the
corresponding BaTiO$_3$ case. Overall, Table III shows similar
displacement patterns for the BaTiO$_3$ and PbTiO$_3$ (011)
surfaces, as well as qualitatively similar results for both {ab
initio} and classical shell-model descriptions.

\subsection{BaTiO$_3$ and PbTiO$_3$ (001) and (011) surface energies}

In the present work, we define the unrelaxed surface energy of a
given surface termination $X$ to be one half of the energy needed
to cleave the crystal rigidly into an unrelaxed surface $X$ and
an unrelaxed surface with the complementary termination $X'$.
For BaTiO$_3$, for example, the unrelaxed surface energies of
the complementary BaO and TiO$_2$-terminated (001) surfaces
are equal, as are those of the TiO and Ba-terminated (011)
surfaces (and similarly for PbTiO$_3$).  The relaxed surface
energy is defined to be the energy of the unrelaxed surface plus
the (negative) surface relaxation energy.  These definitions are
chosen for consistency with Refs.~[\onlinecite{heif2,heifhf}].
Unlike the authors of Refs.~[\onlinecite{bott, heifbz}], we have
made no effort to introduce chemical potentials here, so the
results must be used with caution when addressing questions of
the relative stability of surfaces with different stoichiometry.

With these definitions, and using the 7-layer slab geometries
specified in Sec.~II\,B, the energy of the unrelaxed BaTiO$_3$ (001)
surface is
\begin{eqnarray}
E_{\rm surf}^{\rm unr}(X)={\frac{1}{4}}[E_{\rm slab}^{\rm unr}({\rm BaO})
+E_{\rm slab}^{\rm unr}({\rm TiO}_2)-7E_{\rm bulk}]
\end{eqnarray}
where $X$ = BaO or TiO$_2$ specifies the termination,
$E_{\rm slab}^{\rm unr}$(BaO) and
$E_{\rm slab}^{\rm unr}$(TiO$_2$) are the unrelaxed BaO
and TiO$_2$-terminated slab energies, $E_{\rm bulk}$ is
energy per bulk BaTiO$_3$ unit cell, and the factor of four
comes from the fact that four surfaces are created
by the two cleavages needed to make the two slabs.
The relaxation energy for each termination can be computed from
the corresponding slab alone using
\begin{eqnarray}
\Delta E_{\rm surf}^{\rm rel}(X)={\frac{1}{2}}[E_{\rm slab}(X)
-E_{\rm slab}^{\rm unr}(X)],
\end{eqnarray}
where $E_{\rm slab}(X)$ is a slab energy after relaxation.
The relaxed surface energy is then
\begin{eqnarray}
E_{\rm surf}(X)=E_{\rm surf}^{\rm unr}(X)+\Delta E_{\rm surf}^{\rm rel}(X) .
\end{eqnarray}
Similarly, for the BaTiO$_3$ (011) case, a cleavage on a bulk
BaTiO plane gives rise to the complementary TiO and Ba-terminated
surfaces shown in Fig.~2(c) and (d) respectively.  Thus,
\begin{eqnarray}
E_{\rm surf}^{\rm unr}(X)={\frac{1}{4}}[E_{\rm slab}^{\rm unr}({\rm Ba})+
E_{\rm slab}^{\rm unr}({\rm TiO})-6E_{\rm bulk}],
\end{eqnarray}
where the energy is the same for $X$ = TiO or Ba,
$E_{\rm slab}^{\rm unrel}$(Ba) and
$E_{\rm slab}^{\rm unrel}$(TiO) are energies of the unrelaxed
slabs.  Relaxation energies can again be computed independently
for each slab in a manner similar to Eq.~(2).  Finally,
the (011) surface can also be cleaved to give two identical self-complementary
O-terminated surfaces of the kind shown in Fig.~2(b).
%
\begin{table}[b]
\caption{Calculated surface energies for BaTiO$_3$ and PbTiO$_3$
(001) and (011) surfaces (in eV per surface cell).
`SM' indicates comparative results from
the shell-model calculation of Ref.~[\onlinecite{heif1}].
}
\begin{ruledtabular}
\begin{tabular}{clcc}
Surface & Termination & $E_{\rm surf}$ & $E_{\rm surf}$ (SM) \\
\hline
BaTiO$_3$ (001) & \quad TiO$_2$    & 1.07 & 1.40  \\
                & \quad BaO        & 1.19  & 1.45  \\
BaTiO$_3$ (011) & \quad TiO &  2.04  &2.35   \\
                & \quad Ba  &3.24  & 4.14  \\
                & \quad O  &  1.72 & 1.81  \\
PbTiO$_3$ (001) & \quad TiO$_2$ & 0.74 & \\
                & \quad PbO &0.83 & \\
PbTiO$_3$ (011) & \quad TiO&1.36 &  \\
                & \quad Pb & 2.03 & ~ \\
                & \quad O &1.72& ~ \\
\end{tabular}
\end{ruledtabular}
\end{table}
%
\begin{table*}
\caption{Calculated magnitudes of atomic displacements $D$
(in \AA), effective atomic charges $Q$ (in $e$), and bond 
populations $P$ between metal-oxygen nearest neighbors (in $10^{-3}e$)
for the BaTiO$_3$ and PbTiO$_3$ (001) surfaces. 
} 
\begin{ruledtabular}
\begin{tabular}{cccdcdcdcd}
 & & \multicolumn{4}{c}{BaTiO$_3$ (001) surface} &
     \multicolumn{4}{c}{PbTiO$_3$ (001) surface} \\
Layer & Property&
   Ion & \multicolumn{1}{c}{TiO$_2$-terminated} &
   Ion & \multicolumn{1}{c}{BaO-terminated} &
   Ion & \multicolumn{1}{c}{TiO$_2$-terminated} &
   Ion & \multicolumn{1}{c}{PbO-terminated} \\
\hline 
1& $D$ &Ti& -0.123 & Ba & -0.080 & Ti & -0.111 & Pb & -0.150 \\
 & $Q$ &  &  2.307 &    &  1.752 &    &  2.279 &   &  1.276 \\
 & $P$ &  &  126    &    & -30    &    &  114    &   &  54    \\
 & $D$ &O &-0.014  &O   & -0.025 &O   & 0.012  &O  &-0.012 \\ 
 & $Q$ &  &-1.280  &    & -1.473 &    & -1.184 &   & -1.128 \\
 & $P$ &  & -38    &    &   80    &    &   44    &   &  106  \\
2& $D$ &Ba&  0.101 &Ti  &   0.070&Pb  &  0.209 &Ti &  0.121 \\ 
 & $Q$ &  &  1.767 &    &  2.379 &   &  1.275 &    &  2.331 \\
 & $P$ &  & -30    &    &   88    &   &   8     &    &  80    \\
 & $D$ &O& 0.015   &O& 0.056   &O& 0.050& O&  0.091\\
 & $Q$ &  & -1.343 & &-1.418 & &-1.167& &-1.258 \\
 & $P$ &  &  90     & &-30    & & 80    & &  6   \\
3& $Q$ &Ti& 2.365  &Ba & 1.803  &Ti & 2.335  &Pb & 1.358 \\
 & $P$ &  &  104    & & -36    & &  108    & &  24 \\
 & $Q$ &O & -1.371 &O & -1.417 &O  & -1.207 &O &-1.259 \\
 & $P$ &  & -34    &  &  98     &   &  18     &  &  96 \\
Bulk&$Q$&Ba&  1.797&Ba&  1.797 &Pb & 1.354  &Pb& 1.354 \\       
 & $P$ &   & -34    & & -34    &   &  16     &   &  16 \\
 & $Q$ &O&-1.388  &O  &-1.388  &O  &-1.232  &O  & -1.232 \\
 & $P$ & &  98     &   & 98      &   &  98     &   &  98 \\
 & $Q$ &Ti& 2.367  &Ti&  2.367 &Ti &  2.341 &Ti & 2.341 \\
\end{tabular}
\end{ruledtabular}
\end{table*}
In this case the 7-layer slab has the stoichiometry of three
bulk unit cells, so the relaxed surface energy of the
O-terminated (011) surface is
\begin{eqnarray}
E_{\rm surf}(O)={\frac{1}{2}}[E_{\rm slab}(O) - 3E_{\rm bulk}],
\end{eqnarray}
where $E_{\rm slab}(O)$ is the relaxed energy of the slab having
two O-terminated surfaces.  Everything said here about BaTiO$_3$
surfaces applies in exactly the same way to the corresponding
PbTiO$_3$ surfaces.

The calculated surface energies of the relaxed BaTiO$_3$ (001)
and (011) surfaces are presented in Table IV. In BaTiO$_3$, the
relaxation energies of the TiO$_2$ and BaO-terminated surfaces
($-$0.23 and $-$0.11\,eV respectively) are comparable, leading
to rather similar surface energies.  On the (011) surfaces,
however, the relaxation energies vary more strongly with
termination.  For example, we find a relaxation energy of
$-$2.13\,eV for the TiO-terminated surface, much larger
than $-$0.93\,eV for the Ba-terminated surface.  The relaxation
energy of $-$1.15\,eV for O-terminated surface gives rise to a
relaxed energy of the O-terminated surface (1.72\,eV) that is
much lower than the average of the TiO and Ba-terminated
surfaces (2.64\,eV), indicating that it takes much less energy to
cleave on an O$_2$ plane than on a BaTiO plane.  The shell-model
results of Ref.~[\onlinecite{heif1}] for the BaTiO$_3$ surfaces are
given shown for comparison; the results are qualitatively similar,
but there are some significant quantitative differences, especially
for the Ba-terminated (011) surface.

The corresponding results are also given for the (001) and (011)
surfaces of PbTiO$_3$ in Table IV.  The results for the (001)
surfaces are similar to those for BaTiO$_3$, although the
relaxed surface energies are somewhat lower.  For the case
of the (011) surfaces, however, we find a different pattern than
for BaTiO$_3$.  We find a very large relaxation energy of
$-$1.75\,eV for the TiO-terminated surface, compared with
$-$1.08\,eV for the Pb-terminated surface and $-$1.12\,eV for the
O-terminated surface.  The average energy of the TiO and
Pb-terminated surfaces is now 1.69\,eV,
to be compared with
1.72\,eV for the O-terminated surface, indicating that the
cleavage on a PbTiO or an O$_2$ plane has almost exactly the
same energy cost.

\subsection{BaTiO$_3$ and PbTiO$_3$ (001) and (011) surface charge
distribution and chemical bonding}

To characterize the chemical bonding and covalency effects, we
used a standard Mulliken population analysis for the effective
static atomic charges $Q$ and other local properties of the electronic
structure as described, for example, in Ref.~[\onlinecite{cat,boc}].
The results
are presented in Table V.  Our calculated effective charges for bulk
PbTiO$_3$ are +1.354\,$e$ for the Pb atom, +2.341\,$e$
for the Ti atom, and $-$1.232\,$e$ for the O atom. The bond
population describing the chemical bonding is +98\,m$e$ between
Ti and O atoms, +16\,m$e$ between Pb and O atoms, and +2\,m$e$
between Pb and Ti atoms.  Our calculated
effective charges for bulk BaTiO$_3$ are +1.797\,$e$ for the Ba atom,
+2.367\,$e$ for the Ti atom, and $-$1.388\,$e$ for the O atom
indicate a high degree of BaTiO$_3$ chemical bond covalency.
The bond population between Ti and O atoms in BaTiO$_3$ bulk is
exactly the same as in PbTiO$_3$, while that between
Ba and Ti is slightly negative, suggesting a repulsive interaction
between these atoms in the bulk of the BaTiO$_3$ crystal.

For the TiO$_2$-terminated BaTiO$_3$ and PbTiO$_3$ (001)
surfaces, the major effect observed here is a strengthening
of the Ti-O chemical bond near the BaTiO$_3$ and PbTiO$_3$
(001) surfaces, which was already pronounced for the both
materials in the bulk. Note that the Ti and O effective
charges for
bulk BaTiO$_3$ and PbTiO$_3$ are much smaller than those expected
in an ionic model (+4\,$e$, and $-2\,e$), and that the Ti-O
chemical bonds in bulk BaTiO$_3$ and PbTiO$_3$ are fairly
heavily populated for both materials.

\begin{table}
\caption{The $A$-$B$ bond populations $P$ (in $10^{-3}e$) and
interatomic distances $R$ (in \AA) on (011) surfaces of BaTiO$_3$
and PbTiO$_3$.  Symbols I-IV denote the number of each plane
enumerated from the surface.  The nearest-neighbor Ti-O distance is
2.004\,\AA\ and 1.968\,\AA\ in bulk BaTiO$_3$ and PbTiO$_3$,
respectively.
}
\begin{ruledtabular}
\begin{tabular}{llrrllrr}
\multicolumn{4}{c}{BaTiO$_3$ (011) surface} &
  \multicolumn{4}{c}{PbTiO$_3$ (011) surface} \\
Atom A & Atom B & $P$ ~ & $R$ ~ &
Atom A & Atom B & $P$ ~ & $R$ ~ \\
\hline
\multicolumn{4}{l}{TiO-terminated} &
  \multicolumn{4}{l}{TiO-terminated} \\
\hline
Ti(I)      & O(I) & 130 & 2.047& Ti(I)  & O(I)& 132  & 2.019 \\
           & O(II)& 198 & 1.784&        &O(II)& 196  & 1.766 \\
O(II)     &Ti(III)& 112 & 2.009& O(II)  &Ti(III)& 120 & 1.948 \\
          &Ba(III)& $-$24 & 2.808&        &Pb(III)& 24  & 2.826 \\
          &O(III) & $-$26 & 2.837&        &O(III) &$-$20  & 2.857 \\
Ti(III)   &Ba(III)& $-$2  & 3.471& Ti(III)&Pb(III)& 2   & 3.410 \\
          &O(III) & 118 & 2.004&        & O(III)&108  & 1.975 \\
          &O(IV)  & 96  & 2.004&        & O(IV) & 88  & 1.976 \\
Ba(III)   & O(III)&$-$32  & 2.834& Pb(III)& O(III)& 20  & 2.783 \\
          & O(IV) &$-$38  & 2.816&        & O(IV) & 8   & 2.734 \\
O(III)    & O(IV) &$-$30  & 2.834& O(III) & O(IV) &$-$36  & 2.706 \\
\hline
\multicolumn{4}{l}{Ba-terminated} &
  \multicolumn{4}{l}{Pb-terminated} \\
\hline
Ba(I)     &   O(II)& $-$38& 2.664& Pb(I)  & O(II) & 126 & 2.589 \\
O(II)     & Ba(III)& $-$36& 2.850& O(II)  &Pb(III)& 24  & 2.742 \\
          &Ti(III) & 84 & 2.022&        &Ti(III)& 74  &1.902 \\
          &O(III)  &$-$38 & 2.859&        &O(III) &$-$46  &2.739 \\
Ba(III)   & O(III) &$-$36 & 2.834& Pb(III)&O(III) &$-$10  &2.783 \\
          &O(IV)   &$-$36 & 2.834&        &O(IV)  &36   &2.813 \\
Ti(III)   &O(III)  & 76 & 2.004& Ti(III)&O(III) &62   &1.968 \\
          &Ba(III) &$-$2  & 3.471&        &Pb(III)& 0   &3.408 \\
          &O(IV)   &98  & 2.008&        &O(IV)  & 92  &2.018 \\
O(III)    &O(IV)   &$-$46 & 2.825& O(III) &O(IV)  &$-$46  &2.816 \\
\hline
\multicolumn{4}{l}{O-terminated} &
  \multicolumn{4}{l}{O-terminated} \\
\hline
O(I)      &Ba(II)  &$-$26 &2.697 & O(I)   &Pb(II) & 50  &2.503 \\
          &Ti(II)  &168 &1.722 &        &Ti(II) & 128 &1.694 \\
          &O(II)   &$-$24 &2.801 &        &O(II)  &$-$26  &2.689 \\
Ba(II)    &O(II)   &$-$40 &2.664 & Pb(II) &O(II)  & 78  &2.574 \\
          &Ti(II)  &$-$2  &3.306 &        &Ti(II) & 4   &3.094 \\
Ti(II)    &O(II)   &82  &2.040 &Ti(II)  &O(II)  &92   &1.977 \\
          &O(III)  &112 &1.689 &        &O(III) &126  &1.831 \\
O(II)     &O(III)  &$-$12 &2.825 & O(II)  &O(III) &$-$24  &2.779 \\
Ba(II)    &O(III)  &$-$10 &2.968 &Pb(II)  &O(III) &26   &2.716\\
O(III)    &O(IV)   &$-$14 &2.945 &O(III)  &O(IV)  &$-$42  &2.842 \\
          &Ti(IV)  &60  &2.159 &        &Ti(IV) &60   &2.051 \\
          &Ba(IV)  &$-$24 &2.767 &        &Pb(IV) &$-$2   &2.712 \\
\end{tabular}
\end{ruledtabular}
\end{table}

The Ti-O bond population for the TiO$_2$-terminated BaTiO$_3$
and PbTiO$_3$ (001) surfaces are +126\,m$e$ and +114\,m$e$
respectively, which is about 20\% larger than the relevant value
in the bulk. In contrast, the Pb-O bond population of +54\,m$e$)
is small for the PbO-terminated PbTiO$_3$ (001) surface, and
the Ba-O bond population of $-$30\,m$e$ is even negative for the
BaO-terminated BaTiO$_3$ (001) surface, indicating a repulsive
character. The effect of the difference in the chemical bonding
is also well seen from the Pb and Ba effective charges in the
first surface layer, which are close to the formal ionic charge of
+2\,$e$ only in the case of the BaTiO$_3$ crystal.

The interatomic bond populations for three possible
BaTiO$_3$ and PbTiO$_3$ (011) surface terminations
are given in Table VI. The major effect observed here
is a strong increase of the Ti-O chemical bonding near
the BaTiO$_3$ and PbTiO$_3$ (011) surface as compared
to already large bonding near the (001) surface
(+126\,m$e$ and +114\,m$e$, respectively) and in the bulk
(+98\,m$e$). For the O-terminated (011) surface
the O(I)--Ti(II) bond population is as large as +168\,m$e$
for BaTiO$_3$ and +128\,m$e$ for PbTiO$_3$, i.e.,
considerably larger than in the bulk and on the (001)
surface.

Our calculations demonstrate that for the TiO-terminated
BaTiO$_3$ and PbTiO$_3$ (011) surfaces, the Ti-O bond
populations are larger in the direction perpendicular to
the surface (+198\,m$e$ for BaTiO$_3$ and +196\,m$e$ for PbTiO$_3$)
than in plane
(+130\,m$e$ for BaTiO$_3$  and +132\,m$e$ for PbTiO$_3$).
The Ti-O bond populations
for the TiO-terminated PbTiO$_3$ (011) surface in the direction
perpendicular to the surface is twice as large as the
Ti-O bond population in PbTiO$_3$ bulk.

\begin{table}[t]
\caption{Calculated Mulliken atomic charges $Q$ (in $e$) and
changes in atomic charges $\Delta Q$ with respect to the bulk
charges (in e) on (011) surfaces of BaTiO$_3$
and PbTiO$_3$. for three terminations. The  Mulliken
charges are
2.341\,$e$ for Ti, $-$1.232\,$e$ for O, and 1.354\,$e$ for Pb
in bulk PbTiO$_3$, and
2.367\,$e$ for Ti, $-$1.388\,$e$ for O, and 1.797\,$e$ for Ba
in bulk BaTiO$_3$.
}
\begin{ruledtabular}
\begin{tabular}{lrrlrr}
\multicolumn{3}{c}{BaTiO$_3$ (011) surface} &
  \multicolumn{3}{c}{PbTiO$_3$ (011) surface} \\
Atom & $Q$ & $\Delta Q$ & Atom & $Q$ & $\Delta Q$\\
\hline
\multicolumn{3}{l}{TiO-terminated} &
  \multicolumn{3}{l}{TiO-terminated} \\
\hline
Ti(I)&   2.216&$-$0.151&Ti(I)&   2.212&$-$0.129 \\
O(I)&$-$1.316&   0.072&O(I)&$-$1.261&$-$0.029\\
O(II)&$-$1.155&   0.233&O(II)&$-$1.057&   0.175\\
Ba(III)&   1.757&$-$0.04&Pb(III)&   1.253&$-$0.101\\
Ti(III)&   2.353&$-$0.014&Ti(III)&   2.328&$-$0.013\\
O(III)&$-$1.299&   0.089&O(III)&$-$1.18&   0.052\\
O(IV)&$-$1.402&$-$0.014&O(IV)&$-$1.239&$-$0.007\\
\hline
\multicolumn{3}{l}{Ba$-$terminated} &
  \multicolumn{3}{l}{Pb$-$terminated} \\
\hline
Ba(I)&   1.636&$-$0.161&Pb(I)&   1.122&$-$0.232\\
O(II)&$-$1.483&$-$0.095&O(II)&$-$1.140&   0.092\\
Ba(III)&   1.799&   0.002&Pb(III)&   1.340&$-$0.014\\
Ti(III)&   2.368&   0.001&Ti(III)&   2.343&   0.002\\
O(III)&$-$1.446&$-$0.058&O(III)&$-$1.277&$-$0.045\\
O(IV)&$-$1.392&$-$0.004&O(IV)&$-$1.247&$-$0.015\\
\hline
\multicolumn{3}{l}{O$-$terminated} &
  \multicolumn{3}{l}{O$-$terminated} \\
\hline
O(I)&$-$1.158&   0.23&O(I)&$-$1.011&   0.221\\
Ba(II)&   1.766&$-$0.031&Pb(II)&   1.257&$-$0.097\\
Ti(II)&   2.213&$-$0.154&Ti(II)&   2.237&$-$0.104\\
O(II)&$-$1.452&$-$0.064&O(II)&$-$1.261&$-$0.029\\
O(III)&$-$1.317&   0.071&O(III)&$-$1.215&   0.017\\
Ba(IV)&   1.792&$-$0.005&Pb(IV)&   1.355&   0.001 \\
Ti(IV)&   2.317&$-$0.05&Ti(IV)&   2.317&$-$0.024\\
O(IV)&$-$1.407&$-$0.019&O(IV)&$-$1.233&$-$0.001\\
\end{tabular}
\end{ruledtabular}
\end{table}

In Table VII we present the calculated Mulliken effective charges $Q$,
and their changes $\Delta Q$ with respect to the bulk values,
near the surface.  We analyzed the charge redistribution between different
layers in slabs with all three BaTiO$_3$ and PbTiO$_3$
(011) surface terminations. The charge of the surface Ti
atoms in the TiO-terminated BaTiO$_3$ and PbTiO$_3$ (001)
surface is reduced by 0.151\,$e$ and 0.129\,$e$, respectively.
Metal atoms in the third layer lose much less charge. Except
in the central layer (and, in the case of PbTiO$_3$, in the subsurface
layer), the O ions also reduce their charges, becoming
less negative. The largest charge change is observed
for BaTiO$_3$ and PbTiO$_3$ subsurface O atoms
(+0.233\,$e$ and +0.175\,$e$, respectively). This gives a
large positive change of +0.466\,$e$ and +0.350\,$e$ in the
charge for each BaTiO$_3$ and PbTiO$_3$ subsurface layer.

On the Ba-terminated and Pb-terminated BaTiO$_3$ and
PbTiO$_3$ (011) surface, negative changes in the
charge are observed for all atoms except for Ba and Ti
in the BaTiO$_3$ third layer, Ti atom in the PbTiO$_3$
third layer, and subsurface oxygen atom in PbTiO$_3$.
The largest charge changes are at the surface Ba
and Pb ions. It is interesting
to notice that, due to the tiny difference in the chemical
bonding between BaTiO$_3$ and PbTiO$_3$ perovskites, the
charge change for the BaTiO$_3$ subsurface O ion ($-$0.095 \,$e$)
and PbTiO$_3$ subsurface O ion (+0.092\,$e$) have practically
the same magnitude, but opposite signs.

For the O-terminated BaTiO$_3$ and PbTiO$_3$ (011) surfaces, the
largest calculated changes in the charge are observed for the BaTiO$_3$
and PbTiO$_3$ surface O atom (+0.230\,$e$ and +0.221\,$e$,
respectively). The change of the total charge in the second
layer is negative and almost equal for both materials.
For the BaTiO$_3$ crystal, this reduction by 0.249\,$e$ comes
mostly from Ti atom ($-$0.154\,$e$). In the PbTiO$_3$ crystal,
the reduction by 0.230\,$e$ appears mostly due to a
decrease of the Ti atom charge by 0.104\,$e$, as well as
a reduction of the Pb atom charge by 0.097\,$e$.

\section{Conclusions}

In summary, motivated by the scarcity of experimental investigations
of the BaTiO$_3$ and PbTiO$_3$ surfaces and the contradictory experimental
results obtained for the related SrTiO$_3$ surface
\cite{bic,hik}, we have carried out predictive electronic
structure calculation to investigate the surface atomic and electronic
structure of the BaTiO$_3$ and PbTiO$_3$ (001) and (011) surfaces.
Using a hybrid B3PW approach, we have calculated the surface
relaxation of the two possible terminations (TiO$_2$ and BaO or PbO)
of the BaTiO$_3$ and PbTiO$_3$ (001) surfaces, and three
possible terminations (TiO, Ba or Pb, and O) of the
BaTiO$_3$ and PbTiO$_3$ (011) surfaces.  The data obtained
for the surface structures are in a good agreement with
previous LDA calculations of Padilla and Vanderbilt
\cite{pad}, the LDA plane-wave calculations of
Meyer {\it et al.} \cite{mey}, and in fair
agreement with the shell-model calculations of
Heifets {\it et al.} \cite{heif1}

According to our calculations, atoms of the first surface layer
relax inwards for BaO and PbO terminated (001) surfaces of both
materials. Outward relaxations of all atoms in the second layer
are found at both terminations of BaTiO$_3$ and PbTiO$_3$ (001)
surfaces.  In BaTiO$_3$, the rumpling of the TiO$_2$-terminated
(001) surface is predicted to exceed that of the BaO-terminated
(001) surface by a factor of two.  In contrast, PbTiO$_3$
exhibits practically the same rumplings for both (TiO$_2$ and
PbO) terminations.
Our  calculated surface energies show that
the TiO$_2$-terminated
(001) surface is slightly more stable for both
materials than the BaO or PbO-terminated (001)
surface. The O-terminated BaTiO$_3$ and
TiO-terminated PbTiO$_3$ (011) surfaces have
surface energies close to that of the (001) surface.
Our calculations suggest that the most unfavorable
(011) surfaces are the Ba or Pb-terminated ones
for both the BaTiO$_3$ and PbTiO$_3$ cases.
We found that relaxation of the BaTiO$_3$ and
PbTiO$_3$ surfaces for is considerably stronger
for all three (011) terminations than for the (001) surfaces.
The atomic displacements in the third plane
from the surface for the three terminations
of BaTiO$_3$ and PbTiO$_3$ (011) surfaces are
still large.
Finally, our {\it ab initio} calculations indicate a
considerable increase of Ti-O bond covalency
near the BaTiO$_3$ and PbTiO$_3$ (011)
surface relative to BaTiO$_3$ and PbTiO$_3$
bulk, much larger than for the (001) surface.

\section{Acknowledgments}

The present work was supported by Deutsche Forschungsgemeinschaft (DFG)
and by ONR Grant N00014-05-1-0054.

\end{document}